# Simple model for turbulence intermittencies based on self-avoiding random vortex stretching

Nicolas Rimbert
Nancy University-CNRS
LEMTA, ESSTIN, 2 rue Jean Lamour,
F-54519 Vandoeuvre cedex

**Abstract**

Whether statistics of intermittencies (between small and large eddies) in homogeneous and isotropic turbulence should be described by a log-Poisson, a log-stable probability density function or other, is still debated nowadays. In this paper, a bridge between polymer physics, self-avoiding walk and random vortex stretching is established which may help to obtain new insights on this topic. A very simple relationship between the stability index of the Lévy stable law and Flory's exponent stemming from statistics of linear polymer growth is established. Moreover the scaling of turbulence intermittencies with Reynolds number is also explained and the overall picture is given of smallest vortex tubes of Kolmogorov length width (i.e. the smallest dissipative eddies) bent by bigger vortices of Taylor length scale (i.e. the mean dissipative eddies), themselves stretched by the bigger eddies in a continuous cascade. This results in a both simple and sound model with no fitting parameters required.

## 1 Introduction

Since the discovery of intermittencies (between small and large eddies) in turbulence by Batchelor and Townsend in 1949 (see Frisch [1], for a review including Richardson's first insights concerning turbulence: "*big whorls have little whorls, which feed on their velocity and little whorls have lesser whorls, and so on to viscosity*" a parody of a famous poem by Swift which is the real guideline of this paper), there have been numerous debates on how to model this phenomenon properly. One of the first examples of modelling was proposed by Obukhov and Kolmogorov [2] who used a lognormal law for the distribution of the dissipation of turbulence. Actually Kolmogorov [3], had obtained this law in previous studies on pulverization/fragmentation and Oboukhov [4] used it later in turbulence modelling. The next important step was the definition of "scale similarity" by Novikov and Stewart followed by Yaglom [5, 6]. This led to a lot of interesting works on multifractal modelling of turbulence (again Frisch is a good reference; see also Evertz and Mandelbrot [7] for the definition of a multifractal measure) which ultimately resulted in multiscale analysis (for instance resorting to wavelets theory cf. [8]). However this has not led to any simple law being established. Also, in the wake of Mandelbrot's work in economics (where log stable laws were widely used, a review can be found in ref. [9]), Kida [10,11] generalized the Kolmogorov-Obukhov results to a log-stable law with a stability index $\alpha = 1.65$. In the meantime, log-stable laws had been widely studied in geophysics where their role as "universal multifractal" (they are universal attractor for cascade processes) was emphasized by Schertzer and Lovejoy [12]. Lastly, the idea of a bridge between turbulence, vortex stretching, polymer growth and self-avoiding walk can be found in the works of Chorin (see [13] for instance). However, no simple connection between them is established.

The purpose of this paper is to show that Kida's empirical analysis can be expressed on a more solid basis as long as the value of the stability parameter is changed from 1.65 to $1/\nu$ where $\nu$ stands for the Flory exponent, well known in polymer physics (this leads to the value 1.70). In section 2, a bridge between an effective vortex stretching mechanism introduced by Kuo [14] and (linear) polymer growth is drawn leading to a mapping of one mechanism onto the other. The scaling properties of three dimensional self-avoiding walks are then used to obtain the results. In section 3, we will show how this model can be successfully applied to a given set of experimental and numerical results. Lastly, the number of stretching and bending events is related to the ratio between the Taylor micro scale and the Kolmogorov scale, which leads to a quite simple model, very close to certain insights of Tennekes [15] concerning the worm-like structure of turbulent vortices. In a closely related way, the application of this modelling to the Rice-Liepmann theorem (giving the average distance between stagnation points of the instantaneous velocity field) is then considered.

## 2   Random Vortex Stretching

By partly following Kuo's work, a very simple way of modelling random vortex stretching can be obtained by simplifying the vorticity equation:

$$\frac{\partial \mathbf{\Omega}}{\partial t} + (\mathbf{v}.\nabla)\mathbf{\Omega} = \mathbf{\Omega}\nabla\mathbf{v} + \nu\Delta\mathbf{\Omega} - \mathbf{\Omega}(\nabla.\mathbf{v}) - \tfrac{1}{2}\nabla\frac{1}{\rho}\times\nabla p \tag{1}$$

Where $\mathbf{\Omega}$ is, here, defined as $\tfrac{1}{2}\nabla\times\mathbf{v}$ (so that it exactly represents the antisymmetric part of the rate of strain tensor i.e. the rotation rate of fluid particles). The first term of the right side member stands for vortex stretching, the second term for viscous diffusion, the third term for vorticity increase related to vortex thinning and the last term is the baroclinic creation of vorticity. Neglecting the second and fourth terms (i.e. viscosity effects and baroclinic generation of vorticity) and setting

$$B(t) = \nabla\mathbf{v}.\frac{\mathbf{\Omega}}{\|\mathbf{\Omega}\|} - \nabla.\mathbf{v} \tag{2}$$

which will be hereafter called "effective stretching", one gets

$$\frac{D\mathbf{\Omega}}{Dt} = B(t)\mathbf{\Omega}, \tag{3}$$

where D/Dt stands for the material derivative. According to Kuo, the random nature of effective vortex stretching ensures that the vorticity evolves toward a log-normal distribution. This shall be examined further in the next sections; it will in fact be shown that this hypothesis is not compatible with the topological constraints that affect a vortex line.

### 2.1   Angular acceleration and stretching

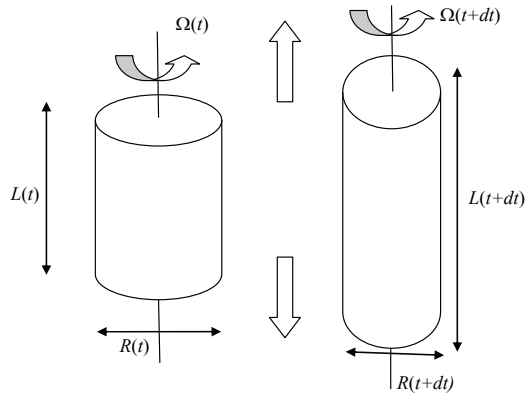

**Fig. 1: axial stretching of a portion of a vortex tube**

Consider a part of a vortex tube, of length $L(t)$, of radius $R(t)$ and angular velocity $\Omega(t)$, submitted to an axial stretching (cf. Fig. 1). In non-viscous flow, vorticity is conserved and materially linked to the fluid particles. Thus the following simple conservation equation can be written:
Mass conservation:

$$R^2 L = \text{constant}. \tag{4}$$

Angular momentum conservation:

$$R^2 \Omega = \text{constant}. \qquad (5)$$

Angular velocity increase can thus be considered proportional to the vortex length increment. It should be noted that this analysis is similar to the pioneering works of Lagrange, Helmholtz, Kelvin, Cauchy,…etc. (see [16], for a more rigorous approach and a review. More details on the present, simpler, analysis can be found in [17]). Using (3) this can be summed up to provide a continuous time stochastic multiplicative process of the form:

$$DL_t = B_L(t) L_t Dt. \qquad (6)$$

If we use the logarithmic derivative, we thus obtain:

$$D \ln L_t = B_L(t) Dt. \qquad (7)$$

The effective stretching $B$ is *a priori* an unknown random variable (and is *a priori* different in (3) and (7) since in this process the vector nature of equation (3) is lost). The next section will determine which scaling constraint the random variables $B$ shall obey but, for the sake of simplicity, continuous time process (7) will be turned into a discrete time process: introducing the logarithmic strain $H = \ln L$, (named $H$ for Hencky since it is very similar to Ludwik-Hencky [18,19,20] "natural" strain used in the modelling of hyper elastic materials) (7) reads:

$$H_{t+dt} - H_t = \ln L_{t+dt} - \ln L_t = B_L(t) dt \qquad (8)$$

Now, let us consider just the time step $(t_1,…,t_n)$ where a "bending event" of a vortex tube does occur. In this case we obtain:

$$H_{t_{n+1}} - H_{t_n} = \int_{t_n}^{t_{n+1}} B_L(t) dt \qquad (9)$$

where the exact properties of the stochastic integral are purposely left undetermined. We shall however assume the independence of the increment of $H$. This is expected to work in the special case of isotropic and homogeneous turbulence. Consequently, it will be supposed that the average helicity is null so that vortex lines do not tend to be aligned in a specific direction. As for the definition of a bending event (cf. Fig. 2), let us define it as the time instant where a new cusp of a given angle (for instance $\pi/2$) appears on the vortex line. Actually the use of the word "cusp" is a loose approximation since it is unlikely that the vortex tube develops an exact singularity.

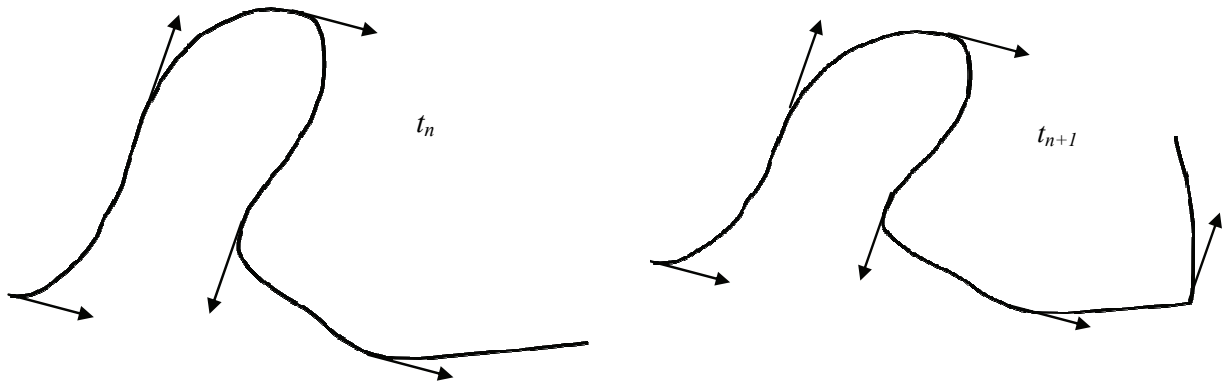

**Fig. 2: turning the continuous cascade of bending events into a discrete one. A bending event is defined by the appearance of a new quasi-cusp on the vortex line: the angle between two successive tangent vectors to the line is equal to a given number (for instance $\pi/2$ in the considered case)**

## 2.2 Topological constraints on a vortex tube and self-avoiding walk

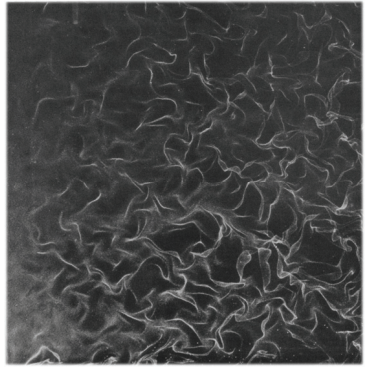

**Fig. 3:** Wrinkling of a fluid surface in isotropic turbulence: Karweit, John Hopkins Univ. 1968 [21]. Here, material lines are stretched, but a clear analogy with vortex tubes can be made, since they can be considered as materially linked to the fluid [16]. (reprinted with permission of the author).

To obtain the simple equation (7), several simplifications and hypotheses were made and particularly we opted to leave out viscosity. This hypothesis holds as long as the finer scale of turbulence known as the Kolmogorov scale $\eta$ has not been reached. At this scale viscosity and inertial term balance each other, so that the former can turn the latter into heat. In this paper the following assumption is made: the fluid possesses a null viscosity until the Kolmogorov scale is reached, where viscosity is applied. Neglecting viscosity ensures that Helmholtz's theorem on vorticity conservation does apply [16]. As a sequel to this theorem, vortex line cannot intersect each other. Therefore a vortex tube shall obey (3) or (7) under the constraint of non intersection.

Though vortex are mainly found in turbulent flows in the form of vortex tubes or "worms" and, sometimes, vortex sheets, the assumption of simple vortex lines will be maintained in the following for the sake of simplicity. However a vortex sheet can be considered as a set of vortex lines in the same way as a material surface can be considered as a set of material lines. Indeed this is what a visualisation by a line of hydrogen bubbles (cf. Fig. 3 and [21]) suggests: each material line seems compelled to self avoidance. Although some material lines are stretched in these experiments, a clear analogy with vortex tubes can be made since the latter can be considered as materially linked to the fluid [16] (moreover hydrogen bubbles have a tendency to migrate toward the low pressure centre of vortices). It should also be noted that since vortex lines are infinitesimally thin, they cannot *stricto sensu* be submitted to stretching and only vortex tubes can be stretched cf. [17]. So, let us look at the ways a vortex tube/line can fold itself under the combined constraint of effective stretching and self avoidance. Leaving aside the dynamics aspect (7) of vortex line folding, it is clear that, using (9) and considering Fig. 2, the kinematics aspects can be fully translated into the world of linear polymer growth or equivalently of self-avoiding walks. In fact a linear polymer chain of $N$ monomers is an archetype of self-avoiding walks. It is a random walk on a lattice where each direction is chosen randomly at each step but with the constraint that it cannot trespass where it has already passed. Fig. 4 gives a sample picture of a bi-dimensional self-avoiding walk or 2D SAW on a square lattice (which will be used for the sake of simplicity though the overall picture is three dimensional).

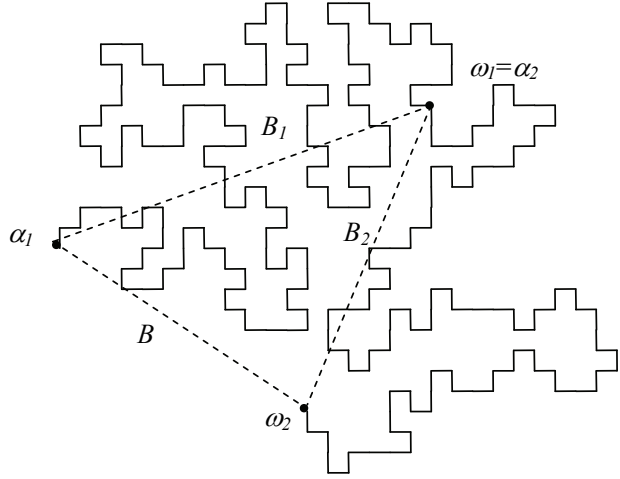

**Fig. 4: Example of a bi-dimensional self-avoiding walk and its decomposition into two sub walks**

According to Flory's pioneering work in polymer physics (see De Gennes [22], for instance), some scaling properties concerning the equivalent radius of a polymer chain of $N$ monomers can be established. Let $N_N(\mathbf{r})$ be the number of chains of $N$ monomers of equal size $a$ going from the origin to point $\mathbf{r}$. Let

$$p_N(\mathbf{r}) = \frac{N_N(\mathbf{r})}{\sum_{\mathbf{r}} N_N(\mathbf{r})} \qquad (10)$$

be the probability distribution of chain of length $N$ and let the gyration radius be defined by:

$$R = \left( \int_0^\infty \mathbf{r}^2 p_N(r) dr \right)^{1/2} \qquad (11)$$

Then it is known that:

$$R \cong aN^\nu \qquad (12)$$

Where $\nu$ is a universal scaling exponent, called Flory's exponent, whose value is appreciatively 3/5 (.588 according to numerical simulations cf. Sokal [23]) in a three dimensional space. For the sake of simplicity we will consider that $a = 1$ from now on. The question which still has to be answered is how the flow maps a piece of vortex tube into a self-avoiding walk similar to that of Fig. 4.

## 2.3 Self-avoiding walk and Lévy-stable stretching

Actually the answer could be quite straightforward namely that the vortex may be strained and bent so as to map onto the self-avoiding walk. But unlike Fig. 4 where each step is equally sized, there is no *a priori* reason that the average step size or standard deviation thereof shall be defined and finite. We shall however use the hypothesis that the average step size is defined and finite (but not necessarily its standard deviation) and we will consider that the actual number of steps is related to the number of bending events. If we accept the hypothesis that each bending event is associated to a stretching event and considering the Hencky strain $H = \ln L$, the latter can be considered to be a good approximation to the number of steps i.e. $N \approx H$. Since the average deviation of the size of steps may eventually not be defined, let us define $\alpha$ as the greatest number so that:

$$\int_0^\infty \mathbf{r}^\alpha p_1(r) dr < \infty \qquad (13)$$

Then the following definition

$$\sigma = \left( \int_0^\infty \mathbf{r}^\alpha p_H(r) dr \right)^{1/\alpha} \quad (14)$$

leads to a finite integral so that (14) can supersede (11) and $\sigma$ can be considered to be a new gyration radius. The preceding hypotheses concerning the finite average size and possibly infinite standard deviation oblige us to consider that $1 < \alpha \leq 2$.

Now, let us look at the scaling properties of this model. If we consider that the number of steps is high enough and that the $H$ step-SAW created by the random effective stretching $B$ can be divided into two SAW, named $B_1$ and $B_2$, each having $H_1$ and $H_2$ steps (cf. Fig. 4) then the following equation is quite obvious:

$$H = H_1 + H_2 \quad (15)$$

And assuming that equation (12) is still verified and takes the shape, $\sigma = \langle H \rangle^\nu$, this leads to:

$$\sigma^{1/\nu} = \sigma_1^{1/\nu} + \sigma_2^{1/\nu} \quad (16)$$

Equation (15) indicates that $H$ shall belong, as a random variable, to the family of Lévy stable law (cf. Feller [24]). To justify this assertion, let us recall that the number of steps is the result of the discrete stochastic process referred in equation (9). As such, the number of steps at a given time is indeed a random variable. From (14), $\sigma$ can thus be considered to be the scaling parameter of the stable law and, using a summing property of these laws, (15) leads to $\sigma^\alpha = \sigma_1^\alpha + \sigma_2^\alpha$ suggesting that the stability index $\alpha$ shall be taken equal to $1/\nu$ so as to recover (16) and more generally (12). Note that the (self-intersecting) case $\nu = 1/2$ leads to Kuo's lognormal model. To conclude this section, the analogy between polymer growth and vortex stretching is summarized in Table 1.

**Table 1: analogy between polymer growth and vortex stretching**

|  | Number of elements | Radius | Relation |
| --- | --- | --- | --- |
| Polymers | $N$ | $R$ | $R = N^\nu$ or $N = R^{1/\nu}$ |
| Vortices | $\langle H \rangle$ | $\sigma$ | $\sigma = \langle H \rangle^{1/\alpha}$ or $\langle H \rangle = \sigma^\alpha$ |

## 3  *Interpretation of some previous experimental and numerical results*

Following Novikov [5,25], the spatial distribution of the rate of turbulent energy dissipation $\varepsilon$ is thought, as $\varepsilon$ represents the death of bigger eddies, to give a proper picture of Richardson turbulent cascade. Actually $\varepsilon$ is never directly measured and, as described by Frisch [1], a bridge between intermittency model based on dissipation and intermittency model based on velocity increment is mostly used. According to Kolmogorov's refined similarity hypothesis the following relation (also known as four-fifth law) can be obtained

$$|u(x+r) - u(x)|^3 = -\frac{4}{5} \varepsilon_r r \quad (17)$$

where $\varepsilon_r$ stands for the dissipation averaged on a sphere of radius $r$. Assuming that the velocity increment and the angular velocity are proportional (*q.v.*), the law of $\ln(|v(r)|)$ or $\ln(\varepsilon)$ shall follow the law of $\ln(|\omega|)$ (or even of $\ln(|\partial u/\partial x|)$) up to a scaling factor and a translation.

In any case, these laws shall be Lévy stable with stable parameter $1/\nu$ and an asymmetry parameter $\beta$, equal to minus one. The last condition stems from a necessary condition for positive moments of the log-Lévy law to be defined and finite.

## 3.1 Comparison to Kida's previous results

Fig. 5 shows the fitting of such a Lévy stable law to the atmospheric turbulence data of Stewart *et al.* [26]. Thought turbulence in the atmospheric boundary layer is not globally isotropic, it is usually considered so, locally, on small scale. The result gives $\alpha = 1.684$ which is closer to $1/\nu$ than the value 1.65 previously used by Kida [11] but obtained without directly fitting the probability density function or PDF (the present fitting procedure is described in [27]). Actually, Kida's value of 1.65 had been previously obtained by fitting the scaling parameter of the moments of the distribution [10] with the data of Anselmet *et al*[28]. The following set of definitions is used in this description: the $p^{th}$ order hyper-flatness of the velocity increment is defined as:

$$H_p(r) = \frac{\left\langle |u(x+r) - u(x)|^p \right\rangle}{\left\langle |u(x+r) - u(x)|^2 \right\rangle^{p/2}} \qquad (18)$$

and the scaling parameter $\xi_p$ as:

$$H_p = \exp(\xi_p) \qquad (19)$$

If $\varepsilon_r$ is log-stably distributed with $\beta = -1$ and using Kolmogorov four-fifths law, this yields

$$\left\langle \varepsilon_r^q \right\rangle = \exp\left[ q m_{\ln\varepsilon} - \frac{\sigma_{\ln\varepsilon}^\alpha}{\cos\left(\frac{\pi\alpha}{2}\right)} q^\alpha \right] \qquad (20)$$

and then

$$\xi_p = -\frac{\sigma_\varepsilon^\alpha}{\cos\left(\frac{\pi\alpha}{2}\right)} \left\{ \left(\frac{p}{3}\right)^\alpha - \frac{p}{2}\left(\frac{2}{3}\right)^\alpha \right\} \qquad (21)$$

Note that the average value $\langle \varepsilon_r \rangle$ depends on all the three values of $m_{\ln\varepsilon}$, $\sigma_{\ln\varepsilon}$ and $\alpha$. This average value is not universal and therefore all these three parameters cannot be universal at the same time. Actually only $\alpha$ is constant in the present modelling and it will be seen in the next paragraph that $\sigma$ does depend on the small scale structure of turbulence (more exactly, on the Taylor and the Kolmogorov scales). It can therefore be inferred that $m_{\ln\varepsilon}$ contains all the large scale dependencies of the dissipation statistics. Fortunately, using definitions (18) and (19) does eliminate this unwanted parameter.

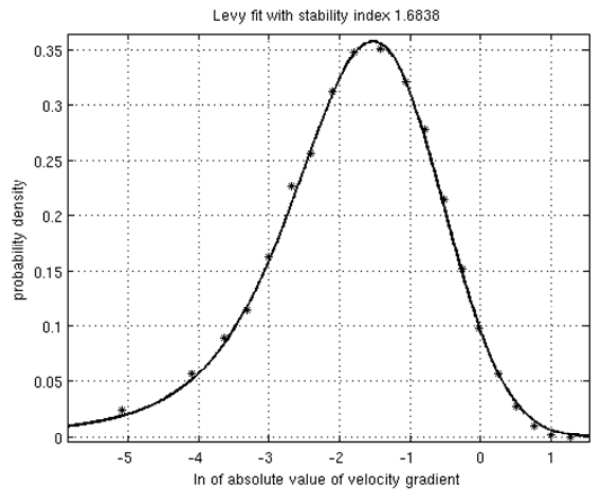

**Fig. 5: Fitting of the PDF of Stewart *et al* [26], with a stable law; Stability parameter is found to be 1.684**

Unfortunately, error bars around values of these moments increase with the order of the moment considered since poorly resolved tails of the distribution are more and more solicited. Moreover knowledge of the PDF is only equivalent to the knowledge of all the moments of the distribution (i.e. an infinite number of moments!). Lastly a distribution having a discrete support (like a Poisson distribution) or a continuous support (like a stable law) may yield similar sets of moments (up to a given order) [29] whereas they are both physically and fundamentally different (due to the topological nature of their support). These arguments illustrate the fact that the moments of a PDF may not be the most appropriate way to describe accurately a PDF and that a direct fitting of the PDF, whenever available, should be preferred. Since these moment transformations are the most common way of post-processing turbulent intermittencies data in scientific literature on the subject, some scaling properties of moments with Reynolds number will be used in the forthcoming section. As well as verifying the present modelling by comparison with new data, it will shed new light on the cascade of stretching events described in previous sections.

### 3.2 Scaling of hyper flatness with Reynolds Number and Tennekes' simple model for the small scale structure of turbulence

Before proceeding, let us discuss the way structure functions can be related to vortex distribution in more detail. Most values of Rimbert and Séro-Guillaume [30] which will be used in the present description have been taken in the near-dissipation range ($\eta < r < 10\eta$) where, according to Chevillard [31] *et al.*, turbulence intermittencies are known to be "rapidly increasing". This may actually be related to the fact that it is precisely the range of scale where the velocity difference is a proper picture of Richardson turbulent cascade. Let us have a look at how, in this modelling, structure function can be computed:

$$\left\langle \left| u(x+r) - u(x) \right|^p \right\rangle = \int \left| u(x+r,\omega) - u(x,\omega) \right|^p d\mathbf{P}(\omega) \qquad (22)$$

where $\omega$ is a random event (i.e. a realization of the vortex tangle) whose probability measure is $\mathbf{P}(\omega)$. Let us assume that point $x$ is in the neighbourhood of a stretched vortex of radius $R$ (actually $R(\omega)$). If $r < R$ then both measurement points in the structure function can be located inside the vortex (let us call this case 1) whereas when $r > R$, at most one point can be located inside the vortex (case 2: one point outside and case 3: two points outside). In the near dissipation range, it will be assumed that viscosity is still non-existent so that the velocity can be assumed to be that of a solid rotation at an angular velocity $\Omega$ inside the vortex. Outside, it is the result of the application of the Biot-Savart law, with the vortices and their intensity superseding electric wires and their current (cf. Lamb [16], chapter VII). Since the geometry of the current is random and somewhat unknown (it is a tangle of random self-avoiding vortices), the velocity outside a vortex can be assumed to be a random variable whose expectation is equal to the mean fluid velocity. Anyway, in case 2 or 3, $u(x+r)$ and $u(x)$ are not much correlated and their contribution to the structure function is hard to figure, whereas in case 1, one gets (appreciatively):

$$u(x+r) - u(x) \cong r\Omega \qquad (23)$$

Actually the hypothesis that the fluid is in solid rotating motion (i.e. a Rankine vortex) is not necessary. Since $r$ is kept fixed, the only thing that matters is that $\Omega$ and velocity increment at fixed $r$, shall be proportional. This means that it still works if the vortex is, for instance, a Lamb or a Burgers [32] vortex. Finally, we obtain:

$$\left\langle \left| u(x+r) - u(x) \right|^p \right\rangle \approx \int r^p \Omega(\omega)^p d\mathbf{P}(\omega) = r^p \int \Omega^p d\mathbf{P}(\Omega) \qquad (24)$$

The structure function can be related to moments of the distribution of vorticity. This can only happen in case 1 which is more frequent in the near-dissipation range. This may explain the so-called increase of turbulent intermittencies in this range. This seems also to agree with the results of Hatakeyama and Kambe [33] for the first, second and third order structure functions: they, for instance, showed that Kolmogorov's four-fifth law is precisely valid in the near-dissipation range.

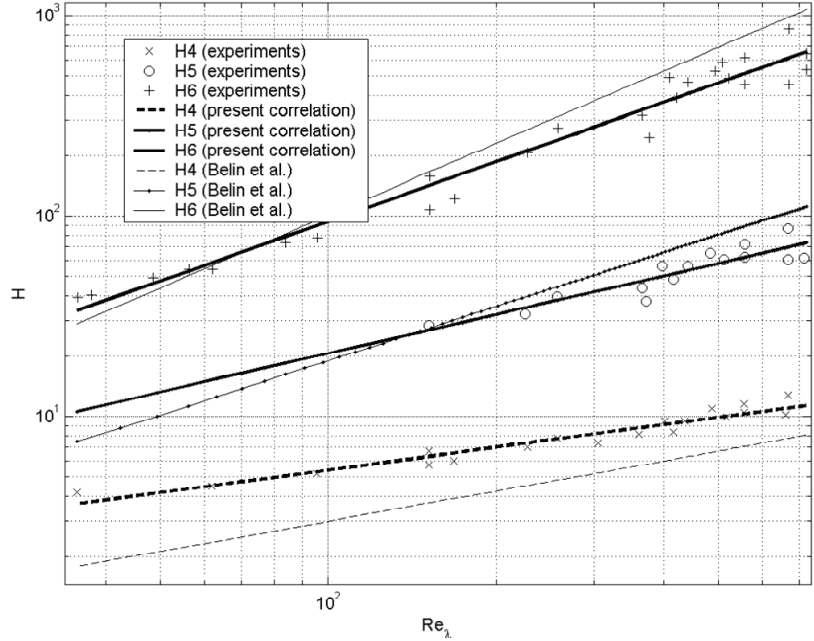

**Fig. 6: scaling of hyper flatness with Reynolds number up to a Taylor scale based Reynolds number of 750 (cf. [30])**

Combining the fitting of the scaling of hyper flatness with Reynolds number of the kind $H_p = a(p).Re_\lambda^{b(p)}$ (cf. Fig. 6, result are summed up in Table 2) and using equations (18-21), this led to the following scaling relation (cf. [30] for details)

$$\sigma_{\ln\varepsilon}^{\alpha} \sim \left[(1,06 \pm 0,08).\ln\left(\sqrt{Re_\lambda}\right)\right] + 0,00 \pm 0,20 \quad (25)$$

common to all the hyper-flatness taken into account (i.e. $p = 4, 5, 6$). In this work a thorough attention to incertitude determination was made (using standard techniques, see Beck and Arnold [34], for instance). Moreover both numerical simulations (Kerr [35]) and experimental results (Belin [36] *et al.*) were used to obtain that correlation. It should be noted that changing the value of the stability index from 1.65 to 1.70 does not change the overall analysis but the results are now on the extreme side of the 95% confidence interval.

**Table 2: scaling of hyper flatness with Reynolds number [30]**

| Order | Hyper-flatness | Moment scaling parameter |
|---|---|---|
| 4 | $H_4 = 0.95\ Re_\lambda^{0,376}$ | $\xi_4 = 0.38\ \ln(Re_\lambda) - 0.045$ |
| 5 | $H_5 = 1.07\ Re_\lambda^{0,642}$ | $\xi_5 = 0.64\ \ln(Re_\lambda) + 0.069$ |
| 6 | $H_6 = 0.99\ Re_\lambda^{0,989}$ | $\xi_6 = 0.99\ \ln(Re_\lambda) - 0.013$ |

It should be remembered that the ratio between Taylor scale $\lambda$ and Kolmogorov scale $\eta$ is equal [37] to:

$$\frac{\lambda}{\eta} = 15^{1/4}.Re_\lambda^{1/2} \quad (26)$$

So that (26) leads to

$$\sigma_{\ln\varepsilon}^{\alpha} \approx \ln\left(\frac{\lambda}{\eta}\right) \quad (27)$$

Where $\lambda$ stands for the Taylor length scale i.e. the mean size of the dissipative eddies. Moreover following section 2.3 and considering the proportionality between vortex stretching *H*, vortex angular velocity $\Omega$ and turbulence dissipation, this equation also reads as follows:

$$\langle H \rangle = O\left(\ln\left(\frac{\lambda}{\eta}\right)\right) \quad (28)$$

This means that the average Hencky strain i.e. the number of stretching and bending events needed to describe the self-avoiding vortex is related to the ratio between the Taylor micro scale and the Kolmogorov scale.

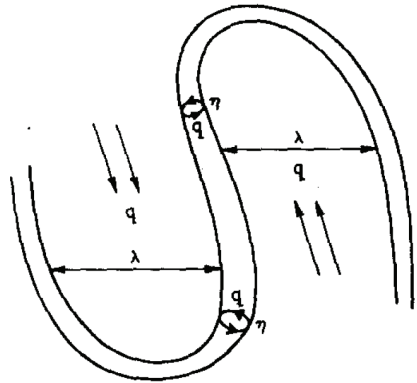

**Fig. 7: Tennekes' simple model for the small scale structure of turbulence [15] (reprinted with permission of the AIP).**

The picture that is suggested can be related to Tennekes' [15] simple model for the small scale structure of turbulence (cf. Fig. 7). A population of vortices of the size of the Taylor micro-scale is both stretching and stretched into a population of vortex tubes whose radii are of the order of the Kolmogorov scale. As Tennekes pointed out by making a budget of turbulent energy dissipation per unit volume (i.e. by multiplying the average dissipation per filament by the filament volume density), the following classic relationship is recovered:

$$\varepsilon = \nu \frac{q^2}{\eta^2} \frac{\eta^2}{\lambda^2} = \nu \frac{q^2}{\lambda^2}, \quad (29)$$

where *q* stands for the scale of velocity fluctuation.

### 3.3 Comparison to other experimental and DNS correlations

The results shown in sections 3.1 and 3.2 mainly come from a previous paper [30] and the main innovation of the present work is the self-voiding random vortex stretching mechanism and the new fitting of data from Stewart et al. [26]. Therefore, it is both interesting and complementary to compare them to different results (without trying to fit the parameters of the law). Unfortunately and it will be explained further in the following, it is very difficult to do so as far as PDF are concerned, since turbulence intermittencies results are mainly published using the moments formalism. A selection of some relevant published results has therefore been made but it is far from being exhaustive.

Firstly, let us compare the present results with the atmospheric data results of Antonia *et al.* [38]. In these, the measured values of the exponent of $H_3$, $H_4$, $H_5$, $H_6$ are 0.11, 0.31, 0.61 and 1.0 whereas the present correlation leads respectively to the values 0.14, 0.35, 0.62 and 0.96. These values come from a strict application of (28), (22) and (20). In [38], confidence intervals are not reported but the present modelling works quite well. The discrepancies increase if the recent wind-tunnel values of 0.09, 0.39, 0.63 and 1.08 obtained by Gylfason *et al.* [39] are considered. But, again, in this experimental study, confidence intervals of the regressions are not reported. It should also be noted that some of these last results have been obtained with an active grid and that the values of $\varepsilon$ range from 0.079 to 7.78 m$^2$/s$^3$. This is somewhat different from DNS data where $\varepsilon$ is usually kept constant and the viscosity is varied so as to increase the Reynolds number. In [40], Ishihara et al. reported DNS values of 0.11 and 0.34 for the exponents of $H_3$ and $H_4$. They also studied the PDF of $|\partial u/\partial x|$, but the values are given between 0 and 30 standard deviations in a linear plot making slightly one decade available for comparison. It should be firstly pointed out that the power-law tail of the Lévy PDF concerns very small fluctuations which are not apparent in linear scale. Secondly, since the greatest reported value of $Re_\lambda$ is 1130, this means that $\sigma_{\ln\varepsilon}^\alpha \approx \frac{1}{2}\ln(1130) \approx 3.5$. Thus the velocity gradient data should be collected on a range of $6\sigma_{\ln\Delta u} = 2\sigma_{\ln\varepsilon} \sim 4.0$ Neperian magnitude scales i.e. 1.7 decades. (Fully asymmetric Levy laws are somewhat different from normal Gaussian laws. It is well known that normal random variables are mainly contained in a ball centred on their average value and of radius equal to 1.96 time their standard deviation i.e. a range of roughly $4\sigma$ whereas that range turns to be closer to $6\sigma$ for fully asymmetric Levy laws. In this reasoning, Kolmogorov's four-fifth law has been used and leads to the following relation between scale parameters of velocity increments and turbulence dissipation: $3\sigma_{\ln\Delta u} = \sigma_{\ln\varepsilon}$. cf. [24, 41]). Actually, as old as it may be, reference [26] may be one of the only references where the post-processing of the PDF of the velocity gradient is done in an appropriate manner as far as comparison with the present modelling is concerned: 6 Neperian magnitude scales are given by Stewart et al. whereas the Reynolds number is estimated to be $10^6$ in [26] so $Re_\lambda \sim 10^3$, $\sigma_{\ln\varepsilon} \sim 2.0$ and 4 Neperian magnitude scales should be roughly expected for the velocity increments. However Stewart et al. gave the statistics of velocity gradients and as pointed out in [42], the scaling used by Stewart et al. was $\varepsilon = (\partial u/\partial x)^2$ therefore the proper relationship is $2\sigma_{\ln(|\partial u/\partial x|)} = \sigma_{\ln\varepsilon}$, which leads to the accurate answer of 6 Neperian magnitude scales.

### 3.4 Application to the Rice-Liepmann theorem

Another interesting upshot of this modelling is that it can be applied to the Rice-Liepmann theorem [43,44,45,46]. This theorem can be related to Tennekes intuitive picture of turbulence summarized in Fig. 7. It states that the average distance $l$ between two zeros of a stationary stochastic function $u$ is equal to:

$$l = C\pi \frac{\langle u^2 \rangle^{1/2}}{\left\langle \left(\frac{du}{dx}\right)^2 \right\rangle^{1/2}} \tag{30}$$

under the assumption that $u$ and $du/dx$ shall be zero-centred and statistically independent. (Application to the non-Gaussian case was suggested by Sreenivasan et al. [47] and the proof is given in [48]). $C$ is the ratio:

$$C = \sqrt{\frac{2}{\pi}} \frac{\left\langle \left(\frac{du}{dx}\right)^2 \right\rangle^{1/2}}{\left\langle \left|\frac{du}{dx}\right|\right\rangle} \tag{31}$$

When $u$ stands for the longitudinal fluctuating velocity, the ratio $\lambda = \langle u^2 \rangle^{1/2} / \left\langle (du/dx)^2 \right\rangle^{1/2}$ is the Taylor micro-scale and the distance $l$ is the average distance between stagnation points of the instantaneous velocity field. Equation (31) now reads $l = C\lambda\pi$. So, as suggested by Fig.7, the average distance between two zeros of the fluctuating velocity field is expected to be proportional to the Taylor micro-scale. *C* equals 1 when *du/dx* is Gaussian and Mazellier and Vassilicos [48] suggest that *C* is actually a function of $Re_\lambda$ and that it should contain some effects of the intermittencies. Let us examine the consequences of the log-stable model. If we use (17) and (20), we thus obtain:

$$\frac{\left\langle\left(\frac{du}{dx}\right)^2\right\rangle^{1/2}}{\left\langle\left|\frac{du}{dx}\right|\right\rangle} = \frac{\left\langle\varepsilon_r^{2/3}\right\rangle^{1/2}}{\left\langle\varepsilon_r^{1/3}\right\rangle} = \exp\frac{\sigma_{\ln\varepsilon}^\alpha}{\cos\left(\frac{\pi\alpha}{2}\right)}\left[\left(\frac{1}{3}\right)^\alpha - \frac{1}{2}\left(\frac{2}{3}\right)^\alpha\right]. \quad (33)$$

which using (25), leads to

$$C = 0.80\,Re_\lambda^{0.06} \quad (34)$$

or $C^{3/2} \cong 0.71 Re_\lambda^{0.09}$. Actually when *a* is small, we get the approximation: $Re_\lambda^a = \exp(a\log Re_\lambda) \approx 1 + a\log Re_\lambda$ and the preceding law can also be written $C^{3/2} \approx 0.71 + .061\log Re_\lambda$ for small $Re_\lambda$. As they expected a dependency of *C* with log ($Re_\lambda$), Mazellier and Vassilicos have reported the following fit: $C^{3/2} \approx 0.87 + 0.11\log Re_\lambda$; for all $Re_\lambda$.. In [48], a collection of experimental results published in scientific literature and of new results is gathered. This leads to a wide range of Reynolds number and gives some statistical significance to the resulting correlation but both active and passive grid turbulence and so-called "chunk" turbulence (Modane wind tunnel) are considered. To circumvent the observed discrepancy, a direct comparison between present (semi-empirical) power-law and data set used in [48] has been made and is presented in Fig. 8. The results of the comparison are quite good, except for data collected in the Modane wind-tunnel experiment. Therefore, with the exception of this data, it seems that the dependence of constant *C* with Reynolds number may be related to some extent to the present model of small scale intermittencies.

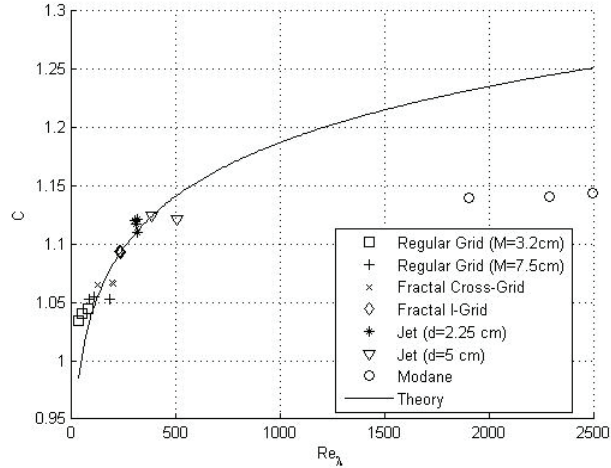

**Fig. 8**: Comparison between predicted curve (34) obtained by application of the Rice-Liepmann theorem to present log-stable law and data collected in [48]. See that reference for a more complete description of the different experimental setups.

## 4   Conclusion

In this paper, a bridge between vortex stretching and linear polymer growth has been drawn leading to log stable law of stability parameter inversely proportional to the Flory exponent. The number of stretching and folding stages needed to map a vortex onto a three dimensional self-avoiding walk has been proven to be related to the logarithm of the ratio between the Taylor micro scale (i.e. the size of the mean dissipative eddies) and the Kolmogorov scale (i.e. the size of the smallest dissipative eddies). Experimental values of the parameters of this modelling have been obtained from four independent sets of experimental and numerical results (PDF from Stewart *et al.* [26]; moments from Anselmet *et al.* [28]; Kerr [35]; Belin *et al.* [36]). All these results merge perfectly, leading to the same value of parameters and giving some generality to the model. Moreover, the resulting model fits reasonably well with other experimental correlations [38, 39] and DNS results [40]. However, post-processing of this data is not usually done in an appropriate manner and therefore a comparison can only be made on a small number of indirect representations of the PDF (e.g. moments). Another interesting result deriving from this model is that it leads to a plausible explanation of the intermittency effect which has been observed when applying the Rice-Liepmann theorem to turbulent flows and to the repartition of stagnation points of the instantaneous velocity field.

There is also an interesting unanswered question remaining namely whether changing definition (11) by definition (14) does influence the value of the scaling exponents $\nu$ and $\alpha$ or not. In fact $\alpha$ should now be properly defined by the following relationship (where long self-avoiding walks of $N$ unitary steps are considered):

$$\alpha / \lim_{N \to +\infty} \tfrac{1}{N} \int_0^{+\infty} |\mathbf{r}|^{\alpha} \, p_N(r) \, dr = 1, \qquad (36)$$

(again the answer would be $\alpha = 2$ in the case of self-intersecting walks). Tests using numerical and statistical sorting [23] of $N$-steps self-avoiding walks could be used to find an answer to this question.

To conclude, however interesting this modelling may be (since it does not contain any free parameter, actually only values of Taylor and Kolmogorov scales do depend on the experiment taken into account), we wish to stress the fact that it can only be considered as a kinematic description insofar as it shows a possible way for the vortices to evolve. The rate and dynamics of vortex evolution is another matter that should be taken into account (this shall be rather simple, in the inertial range, but boundary conditions on the integral scale and Kolmogorov scale are another, tougher, matter). The goal of this work would be, for instance, to explain the relaxation mechanism of homogeneous and isotropic turbulence [49]. Other topics of interest would be the introduction of transitions in the behaviour of vortices (for instance vortex breakdown or reconnections cf. Tabeling and Willaime [50], Chorin [13]) or application to helical flows, all of which may lead to some change in the modelling (maybe in the value of the stability parameter). Lastly links with the pioneering work of Kolmogorov on pulverization (cf. [27]) which was the incentive of this work will be considered in forthcoming papers.